# Tradespace analysis of GNSS Space Segment Architectures

Filipe Pereira and Daniel Selva

*Abstract*— **Global Navigation Satellite Systems (GNSS) provide ubiquitous, continuous and reliable positioning, navigation and timing information around the world. However, GNSS space segment design decisions have been based on the precursor Global Positioning System (GPS), which was designed in the 70s. This paper revisits those early design decisions in view of major technological advancements and new GNSS environmental threats. The rich tradespace between User Navigation Error (UNE) performance and constellation deployment costs is explored and conclusions are complemented by sensitivity analysis and association rule mining. This study finds that constellations at an orbit altitude of ~2 Earth radii can outperform existing GNSS in terms of cost, robustness and UNE. These insights should be taken into account when designing future generations of GNSS.**

## I. INTRODUCTION

THE Global Navigation Satellite System (GNSS) landscape has changed significantly in the last decades. In the 70s, the Global Positioning System (GPS) was designed as a military system to address US Air Force and Navy needs. Later, the system evolved to accommodate civilian needs given that accidents due to human navigation errors were relatively common. At the turn of the century, as GPS and the Russian GLObal Navigation Satellite System (GLONASS) achieved full operational capability, two major events took place: the abolishment of GPS's Selective Availability (SA) and the development of alternative GNSS by other nations. The former enabled precise (m-level) positioning in GPS civilian applications, while the latter allowed the creation of synergies between systems (e.g., shared signal structures in GPS and Galileo (European GNSS)). The current GNSS space infrastructure of nearly 100 satellites integrates a multitude of global constellations and augmentation systems that can be seen as a collaborative system-of-systems. In the next decades, GNSS growth will provide new opportunities (e.g., greater satellite redundancy) but also new challenges (e.g., signal interference [1]).

The importance of GNSS is now well-documented. According to the latest European GNSS Agency (GSA)'s market report [2], there are currently more than 5.8 billion GNSS devices, 80% of which are in smartphones. In 2015 alone, the revenue derived worldwide from GNSS was valued at 95B€ with an expected growth exceeding 10% annually until 2025 [2].

### A. Past GPS architecture decisions

The GPS architecture, which has been used as a case study in the systems architecting literature [3], is an example of reliability and adaptability in space systems design. The success is in part attributed to the significant spin-off value generated (e.g., in civil aviation), which has been far beyond the original plans. The main GPS architecture decisions can be summarized as follows: (1) Use of trilateration in simultaneous one-way range measurements; (2) Passive user operation; (3) System time implementation through synchronized satellite atomic clocks; (4) Low-cost user equipment relying on quartz crystal oscillators; (5) Common transmit signal frequencies using Code Division Multiple Access (CDMA); and (6) 12-h synchronous orbits in Medium Earth Orbit (MEO).

GNSS constellations architected decades later, such as Galileo or Beidou (Chinese GNSS), largely adopted the same fundamental principles and technology as given by the first 5 architecture decisions, but they differ in the last one. Thus, one of the central points in this paper is the analysis of the implications of choosing different satellite orbit characteristics. There are reasons to believe that operational considerations –facilitation of Initial Operational Capability (IOC), performance assessment, simplified bookkeeping, fixed ground station antennas and deployment of ground stations in US territory only- played an important role in the original orbit selection. Thus, the daily repeatability of satellite ground track was seen as a major advantage of 12h synchronous orbits. The following quote in [4] gives support to this view: "*the most compelling reason for choosing a synchronous orbit is that a synchronous system allows regional coverage at minimum cost while allowing gradual extension of coverage to a global system as additional satellites are launched.*". These operational considerations are arguably less important now, given the existence of a global and sophisticated GNSS ground station network.

### B. Motivation for revisiting the architecture of GNSS

Despite the success of the GPS architecture, we argue that it is the right time to analyze the rationale behind the original architecture decisions and question whether or not there are reasons to revisit them, after decades of operations. Specifically, five factors motivating a new architecture study on GNSS were identified.

First, the initial GPS design excluded Low Earth Orbit (LEO) because, at that time, satellites possessed an expected lifetime of only 3-5 years, which for LEO constellations implied ~100 satellites to be launched every year. This result was driven by the requirements for global coverage, satellite visibility to at least 4 satellites from anywhere on the Earth's surface, and good satellite geometric diversity. LEO GNSS architectures need hundreds of satellites to achieve those requirements, in contrast to the 24 satellites envisioned in the original GPS architecture in MEO. However, the operational lifetime of GPS satellites has, on average, more than doubled their designed lifetime. This suggests that GPS III satellites (15-year design lifetime) have the potential to remain operational for 30+ years, thus making lower orbits more feasible.

Second, there is now a growing concern about space debris. The current end-of-life disposal plans of GNSS global constellations make use of graveyard orbits, e.g., at 500km higher than the nominal orbit altitude, for both non-operational satellites and rocket upper stages. Most studies [5] indicate that resonance effects induced by Sun/Moon and J2 secular perturbations will cause long-term growth in orbit eccentricity. The same study concludes: "*These results directly impact the safety of future navigation satellites in the altitude region from 19,000 to 24,000 km*". This is because the disposed satellites are predicted to start crossing the operational orbits in 40 years. Satellites in lower orbits could integrate re-entry disposal strategies benefiting from lower end-of-life $\Delta V$ requirements and improving sustainability in the long-term.

Third, advancements in the miniaturization of space-qualified atomic clocks have the potential to lower the payload power requirements and dry mass of future navigation satellites. Thus, larger GNSS constellations of smaller satellites in lower altitude orbits become more competitive.

Fourth, demand for increased GNSS signal performance in challenging environments, such as in indoor positioning or in the presence of jamming signals has led to an increase in GNSS transmit power levels. As an example, the current GPS III satellites transmit civil signals that are 5x more powerful (~250W) than GPS II-R (~50W). This results in a ~7dB increase in received signal power. On the other hand, the same performance improvement would have been experienced, for example, by placing GPS II-R like satellites in a low MEO orbit, such as the 8330 km orbit altitude considered in this study. Given the cost difference (GPS II-R satellites cost ~50 $M is real terms [6], which is 75% less than a GPS III) and assuming that payload power and satellite lifetime are driving the costs, as suggested in [7], this poses interesting trade-offs in GNSS architecture design.

Finally, increased competition in the launch service providers market has substantially decreased the cost of access to space (e.g. through rocket first-stage reusability) and increased both rocket performance and the number of launch opportunities for smaller satellites. Thus, it is interesting to assess how these developments impact the attractiveness of larger GNSS constellations in terms of total space segment cost and orbit maintenance launch requirements.

Past contributions have analyzed the potential for navigation hosted payloads in proposed LEO broadband mega-constellations (e.g. Iridium NEXT, SpaceX, OneWeb) [8]. LEO constellations are found to benefit from a more benign radiation environment - allowing for Commercial-Off-The-Shelf (COTS) components to be used - and lead to better satellite geometries. The GPS ephemeris message is also shown to be capable of representing LEO orbits, with the adjustment of scale factors on the user side. Other notable contributions are a report on the GALILEO constellation design [9] describing a systematic approach that considers geometry repeatability and constellation stability, as well as a constellation design study proposing a uniformly distributed Flower constellation for GNSS [10]. However, these studies have not explored or suggested alternative GNSS space segment architectures at other orbit altitude regimes. Most GNSS design studies have been directed towards improving the functionality of existing constellations, with no consideration for alternative GNSS orbit characteristics. For example, Fernández studies the enhancement of orbit/clock prediction accuracy and broadcast navigation data by inter-satellite links, assuming the Galileo system architecture as a starting point [11]. Hastings and La Tour study the economic impact of space asset disaggregation assuming the current GPS orbit design [12]. However, to the best of our knowledge, there are no published GNSS architecture studies that include a wide range of orbit altitude regimes, in particular, lower orbits in MEO.

## C. Exploring future GNSS architectures through tradespace exploration

We adopt a design space exploration or "design by shopping" approach [13] to help us understand the complex trade space for future GNSS constellations. In this frame, a large number of design alternatives are generated (e.g., by full factorial enumeration). These design alternatives are parameterized by a set of design variables and evaluated by a model computing a set of performance, cost and/or risk metrics based on the values of those design variables and some endogenous model parameters. Thus, the design process is approached as a shopping experience, during which the decision-maker simultaneously elicits his or her preferences while observing the major feasible alternatives and trade-offs between conflicting objectives. Tradespace exploration has been applied to different kinds of space systems [14], including Earth-observing systems and space communication systems (e.g., NASA's Space Communication and Navigation System or SCaN) [15][16]. However, in the scope of GNSS, the application of this tool has been limited to the study of GPS space asset disaggregation [12]. The goal of the paper is to apply the design by shopping framework to explore the tradespace of possible architectures for future GNSS. We use full factorial enumeration instead of optimization (e.g., [17],[18],[19]) because we prioritize understanding tradespace structure and sensitivities rather than only finding the best possible architecture(s). The main research questions we address in this study are the following:

### D. Paper structure

The rest of the paper has the following structure. Section 2 describes the methods used for the enumeration, evaluation and ranking of GNSS space segment architectures, as well as the methods for sensitivity analysis and data mining. The focus is on the derivation of performance and cost metrics and underlying assumptions taken at each step. Section 3 presents the results of the tradespace exploration, sensitivity analysis and data mining performed on the design decisions and driving parameters. Finally, Section 4 summarizes the main tradeoffs identified at different altitude regimes, discusses the limitations of this study, and suggests avenues for future work.

## II. Methods

In this study, tradespace exploration (a.k.a. tradespace analysis) was used to analyze the cost and performance of candidate architectures under the assumptions presented in this section. The method is based on full factorial enumeration as opposed to optimization since we seek to understand existing trade-offs in a wide range of decision options. Additionally, the analysis does not take the time evolution of these systems into consideration.

Reference architectures are important to help anchor the analysis, validate the model and assess its limitations. The reference architectures for this work are the original GPS and Galileo architecture designs, as shown in Table I –not the current operational configuration of these systems. The choice of Galileo, as opposed to GLONASS or Beidou, is justified by the publicly available satellite data, which was extensively used for modeling purposes (e.g, NAV payload component's power consumption). Furthermore, the reference architecture's position in the cost-performance space helped inform existing trade-offs and opportunities for improvement in next-generation GNSS.

The analysis is focused on a bare-bones navigation Satellite Vehicle (SV) containing a single navigation payload capable of transmitting up to three navigation signals in L1, L2 and L5 frequencies. Such a design is similar to the Galileo Full Operational Capability (FOC) satellite without the Search and Rescue (SAR) payload.

TABLE I
REFERENCE ARCHITECTURES

|  | Orbit Altitude [km] | Orbit inclination [deg] | # Orbit planes | # SV per plane | # SV |
|---|---|---|---|---|---|
| GPS | 20,188 | 56 | 6 | 4 | 24 |
| GAL | 23,229 | 56 | 3 | 9 | 27 |

Given the current technological and programmatic environment:

(1) What are the main trade-offs in GNSS space segment architecture design?

(2) Are there alternative constellation designs that could outperform the existing GNSS?

### A. Defining the architecture space

This study assumes that the most important GNSS space segment architecture decisions are the ones characterizing the satellite constellation and the ones driving the satellite mass (transmit power and satellite lifetime). GNSS requirements for global coverage with a minimum number of satellites are particularly well met by Walker-Delta type constellations [20], which are representative of current GNSS architectures. Thus, to simplify the analysis, the constellation design was restricted to the Walker-Delta pattern even though other designs (e.g. Flower constellations [10],[17]) have the potential to outperform it. The following set of architecture decisions –one option to be chosen per decision- shown in Table II was considered in this study.

Orbit altitude options were chosen to ensure orbit repeatability within 2 weeks –short duration orbit repeatability is desirable for GNSS user performance guarantees and to facilitate ground operations –and to capture existing space architectures that could be used for model validation purposes. Examples of constellations with similar orbit altitude regimes are: Iridium (780km), OneWeb (1200km), O3b (8,062km), GPS (20,188km) and Galileo (23,229km). The L1 (C/A) received signal power of GPS IIR-M/II-F satellites ranges from a minimum of -158.5dBW to a maximum of -153dBW [21], whereas the L5 signal transmitted by GPS II-F/III is expected to reach a maximum received signal strength of -150dBW [22]. Even though a signal strength of -145dBW is not currently available for civilian users, this choice is justified due to the expected improvements in signal anti-jamming capability and indoor performance that would arise from it.

TABLE II
ARCHITECTURE DECISIONS

| Architecture decision | Options considered |
|---|---|
| Orbit altitude [km] | 780, 1250, 8330, 12525, 20188, 23229, 30967 |
| Total number of satellites | 20, 24, 27, 30, 48, 60, 84, 96, 360, 480, 600, 720, 840 |
| Orbit inclination [deg] | 87, 56, 64 |
| Number of orbital planes | 3, 4, 5, 6, 24, 30 |
| Received signal power [dBW] | -155, -150, -145 |
| Transmitted signal frequencies | single (L1); dual (L1, L2); triple (L1, L2, L5) |
| Satellite lifetime [years] | 5,10,15 |

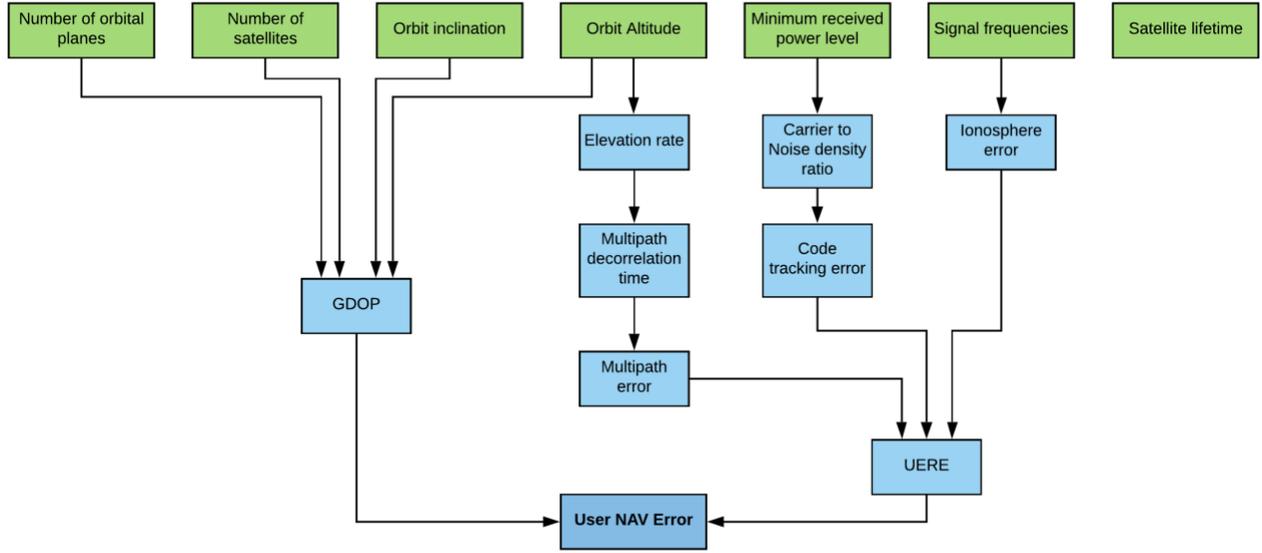

Fig. 1- Architecture design decision to performance metric (UNE) map

The unconstrained full-factorial architectural enumeration of the decisions and options in Table II resulted in 44,226 architectures. Roughly 90% of these architectures were eliminated due to the application of a few constraints. First, 5103 architectures for which the total number of satellites was not an integer multiple of the number of planes were excluded. Second, certain combinations of decision options would produce unreasonable architectures in terms of the cost and performance objectives, i.e., architectures that cannot fulfill the requirements for global navigation service provision due to insufficient coverage – less than 100% - (e.g., 20 satellites at 780 km altitude) or that would result in space segment costs exceeding 60 billion US dollars ($B), i.e. far exceeding those of GPS (e.g., the deployment of an 840 satellite constellation at 30967km altitude). Thus, only architectures consisting of polar orbits with more than 360 satellites and with at least 24 orbital planes in LEO were considered, as well as architectures consisting of non-polar orbits with less than 96 satellites and at most 6 orbital planes in MEO. In this process, 34,101 architectures were eliminated.

For the remaining 5,022 architectures, only those capable of producing acceptable navigation performance, as measured in terms of a maximum Geometric Dilution of Precision (GDOP) value under 6.0 (typical value in GPS performance assessment), were selected. The details of GDOP computation are shown in the next section. The final number of feasible architectures was 4,644.

### B. Evaluating GNSS architectures

The main performance and cost metrics used in this study are: (1) User Navigation Error [m], and (2) Total Space Segment Cost over 30 years [FY2018 $B]. The approach to computing those two metrics is described below.

#### 1) User Navigation Error

The User Navigation Error (UNE) is a scalar combining both position and time error that is derived from the Navigation solution error covariance obtained in the least-squares process, when processing pseudorange measurements. Assuming zero mean and uncorrelated pseudorange measurement errors, UNE is given by

$$UNE = UERE \cdot GDOP \qquad (1)$$

where User Equivalent Range Error (UERE) is a scalar that determines the uncertainty in the ranging signal, and the Geometric Dilution of Precision (GDOP) is a scalar that quantifies user-satellite geometric diversity.

Figure 1 shows the design decisions (shown in green) to performance metric map that provides an overview of the error components contributing to the UNE. These errors are subsequently dealt with in more detail. The assumption was made that in contrast with the cost metric, the satellite lifetime plays no role in the performance metric.

##### a) Geometric Dilution of Precision

For a typical GNSS user located on the earth's surface, GDOP is mainly a function of satellite constellation design parameters. Intuitively, the GDOP value is found to be highly correlated with the volume of a tetrahedron formed by four receiver-satellite unit vectors [23].

To evaluate GDOP, a spherical grid of 41,000 equidistant points was generated, which corresponds to roughly one-degree longitude at the Equator. At each point, GDOP was computed over 1 day for all satellites above 5° elevation - typical in GNSS signal processing. Rigorously, GDOP should be computed for the entire satellite ground track repeatability period. However, we chose 1 day since it was a very good approximation (< 0.02%) of the true GDOP in the test cases and it led to important savings in computational time. Additionally, the time step for DOP computation was chosen to be tenfold the orbit propagation time step (set to correspond to a mean change in true anomaly of 0.5 deg), which resulted in a mean change in satellite elevation of approximately 5 deg. The global average GDOP at the worst site $\langle GDOP \rangle_{worst}$ was determined by computing the average over the simulation period for each unique latitude/longitude point and taking the worst value as the result [24]. An example of how $\langle GDOP \rangle_{worst}$ varies by latitude is shown in Figure 2. The values shown in this figure correspond to the worst (maximum) value of the combined GDOP obtained for all the sites sharing the same latitude. The GDOP variation in longitude is negligible.

All simulations were done with MATLAB and the Orbit Determination Toolbox (ODTBX) [25] using the following force models: Joint Gravity Model 2 with degree and order 20, Sun & Moon Perturbations using JPL DE405 ephemerides, NRL-MSISE2000 global atmospheric model with a drag coefficient of 2.2, and solar radiation pressure with a coefficient of reflectivity of 0.8. Furthermore, a satellite mass of 700kg (from Galileo FOC) and a cross-sectional area of 3m2 was assumed, to compute the results a priori. These assumptions are appropriate -even though these parameters (mass and area) drive the magnitude of drag and solar radiation pressure terms- because the resultant differences in DOP are negligible at the orbit altitudes and time periods under consideration.

### b) User Equivalent Range Error

UERE is typically divided into Signal in Space Ranging Error (SISRE) and User Equivalent Error (UEE). SISRE is largely dominated by the quality of the Orbit & Clock (O&C) determination products, satellite group delay errors and precision of curve fit parameters in the Navigation (NAV) message. The following simplifying assumptions were made in order to restrict the analysis to key parameters in GNSS space segment design: (1) Availability of low latency NAV message dissemination e.g., via inter-satellite links; (2) State of the art on-board atomic clocks, e.g., Passive Hydrogen Maser (PHM) that are able to fit in a smallsat bus; (3) NAV message parameters with appropriate scale and precision for negligible curve fit errors; (4) Availability of GNSS ground segment infrastructure able to provide precise real-time orbit and clock determination products. These assumptions ensure that UERE is not dominated by satellite orbit and clock errors

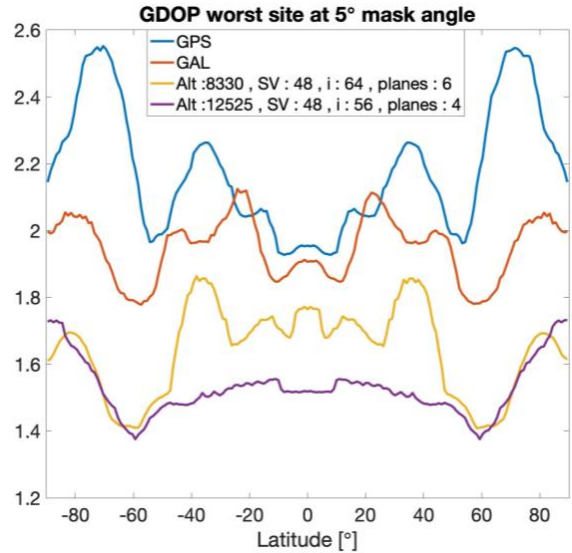

Fig. 2 - Maximum values of the Average Geometric Dilution of Precision (GDOP) over one day at given latitude for GPS, Galileo (GAL) and two proposed architectures

that depend mainly on ground segment infrastructure and capability. Based on the stated assumptions, SISRE is assigned a representative value of 0.5 m, (see ephemeris and satellite clock error budget in [26]).

UEE accounts for the effects of signal propagation and receiver processing errors. The signal propagation error contribution is divided into tropospheric and ionospheric errors. The troposphere is a nondispersive medium for frequencies up to 15 GHz that can be precisely modeled without meteorological information. Assuming the use of the UNB3 tropospheric delay model [27], a constant term of 0.2m is assigned to the tropospheric delay error. The ionosphere is a dispersive medium at GNSS frequencies, which causes an equal amount of signal group delay and carrier phase advance with respect to free space propagation. This phenomenon leads to signal frequency-dependent errors. First-order ionospheric effects can be effectively removed with the iono-free combination in dual-frequency operation [28]. As for single-frequency operation, ionospheric models can be used, such as the Klobuchar model [28], which relies on parameters distributed by the NAV message. The error budget assigned for the ionospheric effect is 4.0 m for single-frequency users and 0.1m for dual and triple-frequency users [26].

As far as receiver processing errors are concerned, multipath and code tracking errors that depend on receiver-satellite dynamics and received signal-to-noise-density ratio (Cs/No) respectively are taken into account. These error sources are especially dominant in demanding environments (such as indoors) when signal atmospheric propagation errors are corrected. Table III shows the code tracking noise standard deviation for the three received signal power options considered in this study. The details regarding the calculation

#### TABLE III
#### CODE TRACKING NOISE STANDARD DEVIATION

| | Received Signal Power $(C_S)[dBW]$ | | |
|---|---|---|---|
| | -155 | -150 | -145 |
| $C_S/N_0\ [dB]$ | 45.9 | 50.9 | 55.9 |
| $\sigma_{tn}\ [m]$ | 0.567 | 0.319 | 0.179 |

of this quantity are shown in Appendix A.

The error due to multipath $\sigma_{mp}[m]$ can be modeled as a function of elevation angle, as suggested in [29], which takes into account the effect of Doppler in receiver signal processing. The elevation angle cut-off is considered to be 5 deg, which constitutes the worst-case scenario in the typical GNSS data processing. Assuming BPSK-R(1) modulation and a receiver front-end bandwidth of 8 MHz, the following expression from [29] was used:

$$\sigma_{mp}[m] = 0.148 + 1.146e^{-0.0471 \cdot Elev} \qquad (2)$$

Multipath is seen as a bias for low user-satellite dynamics, however, as the relative motion increases, it becomes possible to largely eliminate the multipath error given enough time. It can be shown that the approximate multipath decorrelation time, $\tau_{M,D}$ is inversely proportional to the elevation angle rate $\dot{\alpha}$ as shown in [30]. Equation 3 assumes an antenna height of 5 times the wavelength.

$$\tau_{M,D} = \frac{1}{20 \cos(\alpha)\ \dot{\alpha}} \approx \frac{1}{10\dot{\alpha}} [s] \qquad (3)$$

Representative values of median elevation angle rate for each orbit altitude (arch decision #1) were obtained by propagating the satellite orbits and using the same user grid as for GDOP computation. The results are summarized in

. The final 1-sigma multipath error expression depends on the user's willingness to wait, $T_{wait}$ and is given by Equation 4.

#### TABLE IV
#### MULTIPATH SPATIAL DECORRELATION

| | Satellite Orbit altitude [km] | | | | | | |
|---|---|---|---|---|---|---|---|
| | 780 | 1250 | 8330 | 12525 | 20188 | 23229 | 30967 |
| Elev. Rate [mdeg/s] | 66.4 | 59.9 | 14.2 | 9.4 | 6.0 | 5.2 | 3.7 |
| $\tau_{M,D}$ [min] | 1.4 | 1.6 | 6.7 | 10.2 | 15.9 | 18.4 | 25.8 |

If the user is willing to wait longer than the multipath decorrelation time, then the multipath error is assumed to be eliminated. By default, a $T_{wait}$ equal to 1 minute is presumed, which means that the user will experience residual multipath even when considering LEO architectures.

This assumption would be valid for the majority of the GNSS users that use the system for navigation purposes in mobile platforms. Nevertheless, in the sensitivity analysis section, the $T_{wait}$ variable is allowed to vary from 0 to 30 minutes and the impact on the Pareto front architecture results is assessed.

$$\sigma_{multipath}[m] \qquad (4)$$
$$= \begin{cases} 0 & , if\ T_{wait} > \tau_{M,D} \\ \sigma_{mp} \cdot \left(1 - \dfrac{T_{wait}}{\tau_{M,D}}\right), if\ T_{wait} < \tau_{M,D} \end{cases}$$

Finally, the total error budget for single and dual-frequency users is summarized in Table V:

The benefits of adding a third NAV signal frequency have been analyzed in the context of carrier-phase based Precise Point Positioning (PPP). The possibility of performing linear combinations of observations at different frequencies enables fast and reliable resolution of carrier-phase ambiguities at progressively smaller wavelengths. One study suggests that positioning accuracy of triple-frequency PPP can be improved by 19%, 13%, and 21 % compared with the L1/L2-based PPP in the east, north and up directions, respectively [31]. Based on these results, a 17.5% improvement in the total 3D UERE for triple frequency architectures with respect to the dual-frequency values is assumed.

##### 2) Total Space Segment Costs

The feasibility of the proposed GNSS space segment architectures is evaluated in terms of a cost metric taking into account the satellite constellation production costs, $Cost_{prod}$,

#### TABLE V
#### UERE ERROR BUDGET

| | Range Error Standard Deviation [m] | | | |
|---|---|---|---|---|
| | Source | Single freq. | Dual freq. | Triple freq. |
| SISRE | O&C | 0.5 | 0.5 | |
| UEE | Tropo | 0.2 | 0.2 | |
| | Iono | 4.0 | 0.1 | |
| | Code track | [0.18-0.57] | [0.18-0.57] | |
| | Multipath | ≤1.01 | ≤ 1.01 | |
| Total (RMS) | | ≤4.17 | ≤1.18 | ≤ 0.97 |

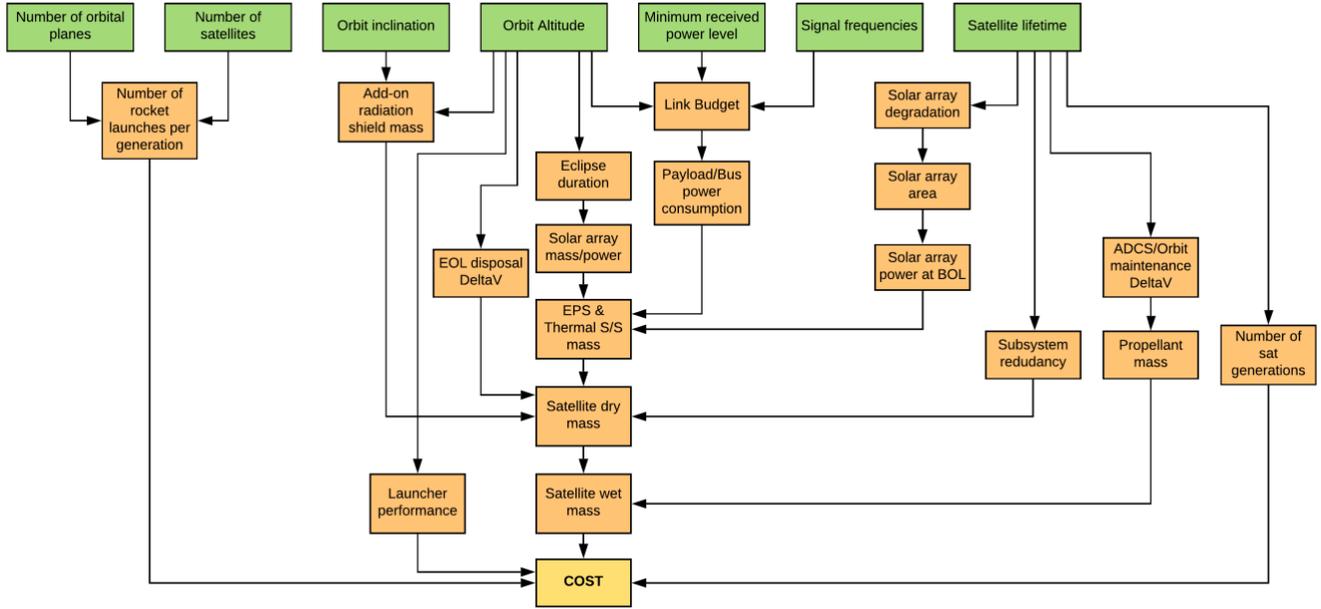

Fig. 3 – Architecture design decision to Cost metric map

as well as the launch costs, $Cost_{launch}$, over a period of 30 years. The time horizon is a multiple of all considered single satellite lifetime options (arch. decision #7) and is long enough to provide the sensitivity of the cost metric to this decision. The impact of operational costs was not considered at this stage, but its contribution is studied in the sensitivity analysis section. Thus, the total space segment costs reported in FY2018 $B is:

$$Cost_{total}\ [\$B] = \frac{Cost_{prod}[\$M] + Cost_{launch}[\$M]}{1000} \quad (5)$$

The design decisions to cost metric mapping is shown in Figure 3 and the derivation of the two cost components is presented below. The satellite mass estimation -driving satellite constellation production costs- is based on first principles and recent spacecraft data to produce an estimate that is representative of the state-of-the-art GNSS satellites. This approach is inherently complex as can be seen in Figure 3. Therefore, to facilitate comprehension, we have moved some of the derivations to the appendix. Also, the structure of the rest of this section follows a top-down approach, starting from the two components of the cost metric (production costs and launch costs) and explaining how they are calculated from its constituents that can be seen in Figure 3.

### a) Satellite Constellation Costs

Satellite production costs were estimated based on the cost of the cheapest commercial platform compatible with the satellite's dry mass, $m_{dry}$, which depends on the values of the architectural decisions. Table VI shows satellite unit costs, $C_{bus}$, for satellite buses that are suitable for the orbit altitude regimes and dry mass values of interest in this study. These values were derived from publicly available data and converted to FY2018 U.S. dollars, considering inflation and euro-to-dollar conversions, when appropriate. The satellite dry mass values in Table VI are assumed to be the maximum allowed at the correspondent price point.

The total satellite constellation cost is given by Equation 6, where $C_{bus}$ is the cost of the cheapest bus option that can host the estimated satellite dry mass

$$Cost_{prod} = C_{bus}N^{(1+\log S/\log 2)} \quad (6)$$

where $S$ is the learning factor (set to 85%), which aims to capture productivity gains as similar flight units are produced and N is the number of flight units required over the 30 years. N is computed taking into account the number of satellites ($N_{sats}$) in the constellation (Arch. decision #1) and the satellite lifetime ($S_{life}$) (Arch. decision #7):

$$N = N_{sats} \cdot \frac{30}{S_{life}} \quad (7)$$

### (1) Satellite dry mass

The satellite dry mass is derived from the estimated satellite power consumption at the end of life (EOL), which depends on the orbit altitude $h_s$ (arch decision #1), number of transmit signals at different carrier frequencies, $N_{freq}$ (arch decision #6) and the target received signal power, $P_R$ (arch decision #5). The payload power requirements are driven by the satellite transmit power that is derived from the link budget analysis in appendix C. The final satellite dry and wet mass estimates were computed as follows:

TABLE VI
REFERENCE DATA FOR SATELLITE UNIT COST ESTIMATION

| Sat. mission | Sat. bus | Orbit alt. [km] | Sat. dry mass [Kg] | Sat. unit cost $C_{bus}$ [$M] | Ref. |
|---|---|---|---|---|---|
| Globalstar | ELiTe 1000 | 1,414 | 350 | 23.11 | [32] |
| GIOVE-A | SSTL 600 | 23,222 | 540 | 39.90 | [33] |
| Galileo FOC | OHB Smart MEO | 23,222 | 660 | 45.70 | [34] |
| O3B | ELiTe 1000 | 8,063 | 800 | 49.72 | [35] |
| GPS IIF | AS-4000 | 20,188 | 1453 | 59.94 | [36] |
| GPS III | A2100 | 20,188 | 2269 | 209.47 | [37] |

$$m_{dry,f}[kg] = m_{dry,i} + m_{RAD} \qquad (8)$$

$$m_{wet}[kg] = m_{dry,f} + m_{propellant} \qquad (9)$$

where $m_{RAD}$ is the mass penalty resulting from the required radiation add-on shielding, $m_{dry,i}$ is the initial satellite dry mass based on the power, thermal, and propulsion subsystems mass estimation and $m_{propellant}$ is the propellant mass required to meet the $\Delta V$ requirements.

According to available Galileo data, the NAV payload mass, $m_{payload}$ is estimated to be approximately 150 kg [38]. This is consistent with subsystem mass distributions from historical spacecraft data [39], as shown in Table VII (Galileo FOC dry mass $\cong 670kg$), which is used in the derivation of the mass quantities for the remaining subsystems: Telemetry, Tracking, and Command (TT&C), Attitude Determination and Control Systems (ADCS) and Structure.

Thus, the following expression for the initial satellite dry mass is obtained:

$$m_{dry,i}[kg] = m_{EPS} + m_{thermal} + n \cdot m_{TT\&C} + \qquad (10)$$
$$m_{ADCS} + m_{propulsion} + m_{struct} + m_{payload}$$

$$= \frac{m_{EPS} + m_{thermal} + m_{propulsion} + m_{payload}}{(1 - n \cdot 0.05 - 0.06 - 0.23)}$$

where $n$ is the number of redundant TT&C elements computed according to the expression suggested in [40], assuming a reference reliability $R_{ref} = 88\%$ at a reference lifetime $T_{ref} = 12\ years$ (from Galileo FOC [41]):

$$n = \frac{\log(1 - R_{ref})}{\log\left(1 - R_{ref}^{S_{life}/T_{ref}}\right)} \qquad (11)$$

Note that $n$ is not necessarily an integer. Its value is meant to capture the dependency of the level of redundancy in the TT&C subsystem with satellite lifetime.

*(a)    Electrical Power and Thermal Subsystem mass estimation*

To assess the Electrical Power Subsystem (EPS) mass, a good estimate of the bus/payload power consumption is necessary. For this purpose, publicly available data applicable to the Galileo FOC satellite was used, which constitutes a good example of a navigation satellite architecture given the absence of other NAV unrelated payloads –in contrast with a modern GPS satellite, which has 7 different payloads. Based on the block diagram produced by the satellite manufacturer [41] and ignoring the Search And Rescue (SAR) functionality, the following core components shown in Table VIII were identified. Furthermore, the RF signal amplification is computed assuming a separate Travelling Wave Tube Amplifier (TWTA) per signal frequency with an efficiency of 68% [42]. Data from the referenced sources were used to produce the payload power budget.

The maximum power consumption value for each component is used to obtain a conservative estimate of the total payload power consumption:

$$P_{PL}[W] = P_{TWTA} + P_{PHM} + P_{RAFS} + P_{FGUU} \qquad (12)$$
$$+ P_{NSGU} + P_{RTU} + P_{Th}$$
$$= (P_T/0.68 + 255)/(1 - 0.15)$$

TABLE VII
NAVIGATION SPACECRAFT SUBSYSTEM MASS DISTRIBUTION (FROM [39])

| S/S | EPS | Payload | Struct. | ADCS | TT&C | Prop. | Thermal |
|---|---|---|---|---|---|---|---|
| | Percentage of satellite dry mass (standard deviation) | | | | | | |
| NAV | 32 (3) | 21 (2) | 23 (3) | 6 (0.5) | 5 (1) | 3 (0.5) | 10 (1) |

TABLE VIII
PAYLOAD POWER CONSUMPTION

| Payload component | Units [#] | Maximum power consumption [W] | Ref. |
|---|---|---|---|
| Phase Hydrogen Maser (PHM) atomic clock, $P_{PHM}$ | 2 | 54 | [43] |
| Rubidium Atomic Frequency Standard (RAFS), $P_{RAFS}$ | 2 | 39 | [44] |
| TWTA amplifier, $P_{TWTA}$ | 1 | $\dfrac{P_T}{0.68}$ | [42] |
| Navigation Signal Generation Unit (NSGU), $P_{NSGU}$ | 1 | 35 | [45] |
| Frequency Generation and Upconversion Unit (FGUU), $P_{FGUU}$ | 1 | 22 | [46] |
| Remote Terminal Unit (RTU), $P_{RTU}$ | 1 | 12 | [47] |
| Payload thermal subsystem, $P_{Th}$ | 1 | 15% of the total payload power consumption | |

where $P_T$ is the satellite transmit power, as computed by the link budget equations in appendix C.

Finally, by analogy with the Galileo FOC satellite, the total satellite bus power consumption is assumed to be 40% higher than the payload power consumption:

$$P_{SC} = 1.4 \, P_{PL} \qquad (13)$$

Once the spacecraft power requirement is defined, the Electrical Power Subsystem (EPS) mass can be estimated by first computing the mass of the solar arrays and batteries. These parameters depend on the orbit altitude, $h_s$ (arch. decision #1) that directly impacts the frequency and duration of solar eclipse events, as well as on the satellite lifetime (arch. decision #7) given the expected degradation of the solar arrays with time. The overall mass of the EPS subsystem is estimated based on a simple mass estimate relationship as suggested in [40], which is considered adequate for this study :

$$m_{EPS} = m_{SA} + m_{bat} + m_{PCU} + m_{dist} \qquad (14)$$

where the mass of the Power Control Unit (PCU) is:

$$m_{PCU} = 0.0045 \cdot P_{BOL} \qquad (15)$$

and the mass of the power distribution system is:

$$m_{dist} = 0.15 \cdot m_{EPS} \qquad (16)$$

To estimate the solar array mass, $m_{SA}$ and the power production at beginning-of-life (BOL), $P_{BOL}$ the steps described in [39] and shown in Appendix B were followed.

The battery mass is computed based on the battery capacity expression assuming space-qualified Li-Ion batteries with an energy density, $u_{bat}[Wh/kg] = 130$ (e.g., Galileo's FOC batteries: ABSL18650HC), a transmission efficiency, $\mu_{trans} = 0.9$ and an average Depth of Discharge (DOD) equal to 30% [39].

$$m_{bat}[kg] = P_{SC} \cdot T_e / (3600 \cdot DOD \cdot \mu_{trans} \cdot u_{bat}) \qquad (17)$$

Similarly, the thermal subsystem mass is estimated as a function of $P_{BOL}$, as suggested in [40]:

$$m_{thermal} = k_P \cdot P_{BOL} \quad , k_P = 0.020 \, kg/W \qquad (18)$$

### (b) Propulsion Subsystem mass estimation

The propulsion subsystem mass was derived using Equation 19 [40]:

$$m_{propulsion}[kg] = 4 + 0.3 \cdot m_{propellant}^{2/3} \qquad (19)$$

where the propellant mass was obtained from the rocket equation assuming Hydrazine monopropellant with specific impulse ($I_{sp} = 227s$), the required $\Delta V$ budget and an initial satellite dry mass estimate dependent on the payload power as derived in [48] and shown in Equation 20. This empirical formula was obtained by analyzing FCC filling data from non-geosynchronous communication satellites that were considered similar to GNSS satellites.

$$m_{dry}[kg] \cong 7.5 \cdot P_{PL}^{0.65} \qquad (20)$$

$$m_{propellant}[kg] = m_{dry}(e^{\Delta V/(9.8 \cdot I_{sp})} - 1) \qquad (21)$$

The required $\Delta V$ budget is computed taking into account satellite orbit correction maneuvers, ADCS, and End-of-Life (EOL) disposal.

$$\Delta V[m/s] = \Delta V_{disp} + \Delta V_{man} + \Delta V_{ADCS} \qquad (22)$$

The $\Delta V$ required for ADCS is estimated based on the values reported by the Galileo FOC Flight dynamics team for propellant mass consumption in Sun Acquisition Mode (SAM) of $\sim 5 g/day$ or $\sim 1.8 \, kg/year$ [49]. Using the Galileo FOC design satellite lifetime ($S_{life} = 12$ years) and satellite dry mass ($m_0 = 700 kg$), and assuming hydrazine monopropellant the resultant $\Delta V_{ADCS}$ was determined with the rocket equation:

$$\Delta V_{ADCS}[m/s] = I_{sp} \cdot 9.8 \cdot \ln\left(\frac{(m_0 + 1.8 * S_{life})}{m_0}\right) \qquad (23)$$
$$\cong 67.6$$

Orbit correction maneuvers are required for the conservation of the overall satellite constellation geometry. Using $\Delta V$ data from AIUB CODE [50] and GPS outage information retrieved from Notice Advisory for Navigation

Users (NANU) messages, as shown in Table IX, the required maneuver $\Delta V$ is estimated to be 0.17m/s every ~500 days. The commensurability of the GPS orbit period with the Earth's rotation period means that the GPS satellites are in 2:1 resonance with geopotential terms and particularly subject to resonance perturbations. Thus, the frequency of station-keeping maneuvers is particularly high for GPS and can be regarded as a worst-case scenario in MEO. In LEO, satellites are subject to stronger gravity perturbations and drag that require frequent station-keeping maneuvers. Considering a ballistic coefficient of approximately 100kg/m2(consistent with our previous assumptions on spacecraft parameters for orbit propagation) the $\Delta V$ for altitude maintenance is estimated to be 0.6 m/s per year @ 780km and 0.08 m/s per year @ 1250km under maximum solar activity conditions [51]. The final expression accounting for the satellite lifetime is the following:

$$\Delta V_{man}[m/s] = \begin{cases} S_{life} \cdot 0.6, \, if \, h_s = 780km \\ S_{life} \cdot 0.08, \, if \, h_s = 1250km \\ S_{life} \cdot 365 \cdot 0.17/500, \, if \, MEO \end{cases} \quad (24)$$

EOL orbital disposal $\Delta V$ requirements are computed assuming a Business as Usual (BAU) strategy per default. For the MEO constellations, this results in the use of a graveyard orbit located at an altitude, $h_e$, 500km higher than the operational altitude: $h_e[km] = h_s + 500$, with a required EOL orbital disposal $\Delta V$ given by [51]:

$$\Delta V_{disp}[m/s] = \sqrt{\mu} \left[ \left[ \sqrt{\frac{2}{r_A} - \frac{2}{(r_A + r_B)}} - \sqrt{\frac{1}{r_A}} \right] + \left[ \sqrt{\frac{1}{r_B}} - \sqrt{\frac{2}{r_B} - \frac{2}{(r_A + r_B)}} \right] \right] \quad (25)$$

where $r_A = r_e + h_s$ and $r_B = r_e + h_e$.

In LEO we assume that the satellite's perigee is reduced to 500km altitude, which ensures a re-entry in less than 25 years assuming typical GNSS mass and area to mass ratios. In this case, the required EOL orbital disposal $\Delta V$ is obtained by Equation 26.

In the sensitivity analysis, the use of a deorbit subsystem that ensures deorbit through atmospheric re-entry at EOL is considered. The required $\Delta V$, in this case, is computed similarly with Equation 26 assuming $h_e[km] = 500$:

$$\Delta V_{disp}[m/s] = \sqrt{\frac{\mu}{r_A}} \left[ 1 - \sqrt{\frac{2r_B}{(r_A + r_B)}} \right] \quad (26)$$

### (c) Radiation Environment

The radiation environment where existing navigation satellites operate, i.e., MEO, is very hazardous when compared

| YEAR | DOY | GPSWEEK | $\Delta V$(mm/s) | Duration (days) |
|---|---|---|---|---|
| SVN45 (PRN21) | | | | |
| 2009 | 205 | 1541 | 129.8 | -- |
| 2011 | 46 | 1623 | 155.2 | 571 |
| 2012 | 241 | 1703 | 173.4 | 560 |
| 2014 | 39 | 1778 | 182.7 | 528 |
| 2015 | 156 | 1847 | 161.5 | 482 |
| 2016 | 267 | 1915 | 212.1 | 476 |
| 2017 | 349 | 1979 | 191.8 | 447 |
| Mean | | | 172.4 | 511 |

to LEO or GEO orbits. It is important to estimate the total radiation dose and add-on shielding required at the given orbit altitude and inclination. Assuming a maximum radiation dose in Silicon of 30kRad at the center of an Aluminum (Al) sphere – appropriate for "careful COTS" components- the necessary Al shielding thickness was determined using ESA's Space ENVironment Information System (SPENVIS) database.

To compute a mass penalty due to add-on shielding, the corresponding satellite volume for our satellite dry mass estimates was computed, assuming a bulk satellite density of $221 \, kg/m^3$ (derived from Galileo FOC volume and mass data). The Al thickness values were added to the radius of a sphere of equivalent volume and the volume of the resultant spherical shell was then converted to a mass penalty, $m_{RAD}$, using a mean Al density of 2700 kg/m3. Radiation sources and effects considered in the simulation are shown in Table X. The IRENE-AE9/AP-9 models contain new data from NASA's Van Allen Probes that are crucial for the correct characterization of the MEO radiation environment. The required Al sphere thickness obtained for the reference GPS and Galileo architectures, as shown in Table XI, was 8 and 7 mm respectively.

| Radiation source | Model parameters |
|---|---|
| Trapped proton and electron fluxes | Proton model AP-9, solar minimum<br>Electron model AE-9, solar maximum |
| Long-term solar particle fluences | ESP-PSYCHIC (total fluence) Ion: H to H<br>Confidence level: 80% |
| Galactic cosmic ray fluxes | Ion range: H to U<br>Magnetic shielding: default |
| Ionizing dose for simple geometries | SHIELDOSE-2 model<br>Center of Al spheres<br>Silicon target |

TABLE XI
REQUIRED ALUMINUM THICKNESS FOR CAREFUL COTS COMPONENTS

| Orbit Altitude [km] | Orbit Inclination [°] | Satellite lifetime | | |
|---|---|---|---|---|
| | | 5 | 10 | 15 |
| | | Aluminum Thickness [mm] | | |
| 780 | 87 | 3 | 4 | 5 |
| 1250 | 87 | 14 | 30 | 40 |
| 8330 | 56 | 8 | 12 | 16 |
| | 64 | 8 | 12 | 14 |
| 12525 | 56 | 7 | 9 | 10 |
| | 64 | 7 | 9 | 10 |
| 20188 | 56 | 7 | 9 | 9 |
| | 64 | 7 | 8 | 9 |
| 23229 | 56 | 7 | 8 | 9 |
| | 64 | 7 | 8 | 9 |
| 30967 | 56 | 6 | 6 | 7 |
| | 64 | 6 | 6 | 7 |

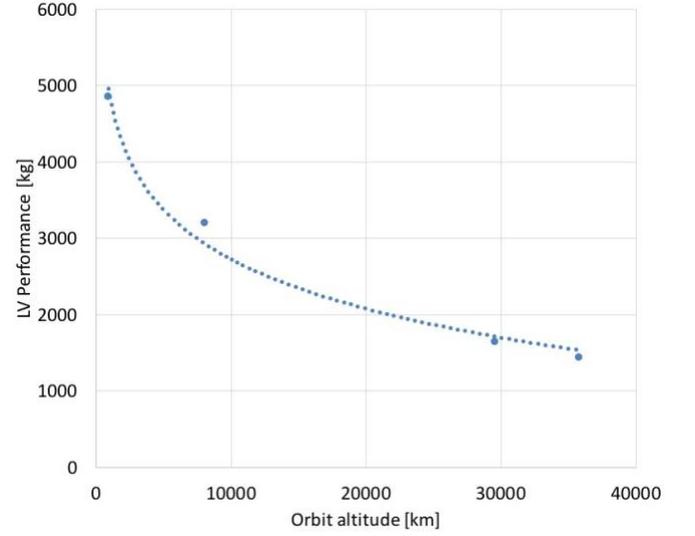

Fig. 4 - Soyuz 2-1B performance data

### b) Launch Costs

The total constellation launch costs, $Cost_{launch}^{const}$, were calculated by determining the minimum number of rocket launches needed for full constellation deployment over a time horizon of 30 years. The primary rocket/upper stage configuration chosen for this analysis was the Soyuz-2/Fregat, given that it has been extensively used for navigation satellite orbit insertions, both in the GLONASS and Galileo programs. Based on data from [52], a fixed cost per launch: $Cost_{launch}^{unit} = 48.5 \ \$M$ is considered. In cases where the satellite wet mass exceeds the Soyuz-2 performance to the desired orbit altitude (some of the constellations at 23,229km and 30,967km), we assume the use of SpaceX's Falcon 9 with a fixed cost per kg, $Cost_{kg}^{unit} = \$25,000$. This value was derived from the recent launch cost of GPS III satellite (wet mass= 3880kg) valued at 96.5 \$M [53]. This constitutes the BAU case, which does not consider first-stage rocket reusability. According to SpaceX, tenfold reusability of the first stage could be achieved resulting in a cost decrease of up to 40% [54]. The impact of lower launch costs was considered in the sensitivity analysis section. The total constellation launch cost was determined by Equation 27, which depends on the minimum number of rocket launches, $N_{min}$ derived from the satellite wet mass estimates and the Soyuz-2 rocket performance, as shown in the next section.

$$Cost_{launch}[\$M] \qquad (27)$$
$$= \begin{cases} Cost_{launch}^{unit} \cdot N_{min}, if \ m_{wet} < perf_{Soyuz} \\ Cost_{kg}^{unit} \cdot m_{wet}.N, if \ m_{wet} > perf_{Soyuz} \end{cases}$$

#### (1) Rocket Performance

Rocket performance is the maximum payload mass that a rocket can safely deploy at a certain orbit altitude. This study uses Soyuz 2-1B rocket data (shown as larger dots in Figure 4) extracted from the user manual [55] and other sources [56] is used to calculate its LV performance, $perf[kg]$, for all orbit altitude options, $h_s \ [km]$, by means of a logarithmic regression model.

$$perf[kg] = -934.4 \cdot \ln(h_s) + 11,333 \qquad (28)$$

Given the satellite dry mass and rocket launcher performance, the maximum number of satellites per launch vehicle was computed, ignoring the additional mass needed for payload adapters in multi-payload launches:

$$N_{sats}^{LV} = floor\left(\frac{perf(h_s)}{m_{wet}}\right) \qquad (29)$$

Furthermore, we added the constraint of a single orbit plane per launch vehicle, given that maneuvers other than orbit phasing require a large amount of $\Delta V$. The resulting expression is given in Equation 30:

$$N_{min} = ceil\left(\frac{N_{sats}^{plane}}{N_{sats}^{LV}}\right) \cdot N^{planes}.N_{sats}^{gen} \qquad (30)$$

where $N_{sats}^{gen}$ is the number of satellite generations in 30 years, $N^{planes}$ is the number of orbital planes (arch decision #4), $N_{sats}^{plane}$ is the number of satellites per plane and $N_{sats}^{LV}$ is the number of satellites per launch vehicle.

### C. Tradespace analysis

#### 1) Pareto analysis

Upon completion of the constrained full-factorial architectural evaluation, the two metrics or objective variables, $o_i(i = 1,2)$ were normalized according to the expression below, where the min/max are with respect to the entire population of architectures:

$$o_i^{norm} = \frac{o_i - min(o_i)}{max(o_i) - min(o_i)} \quad (31)$$

and sorted by Pareto ranking. The sorting algorithm finds architectures that are better than all alternatives in at least one objective (e.g., cost), and no worse in the other objective (e.g., performance). These Pareto optimal solutions are the set of non-dominated architectures and are given a rank of 1 by a process referred to as Pareto ranking. The process continues by iteratively finding the nondominated front among the unranked architectures and incrementing the rank by 1. Thus, the optimal trade-offs between the performance and cost metrics were found.

### 2) Data mining

A reasonable question to answer in tradespace analysis is whether non-dominated architectures have some features in common, such as a specific orbit altitude. This can be done manually using visualization tools (e.g., by creating cost-performance scatter plots where architectures are color-coded according to some decision). To answer this question more systematically, a simple data mining method integrating classification and association rule mining [57] was used. The algorithm can identify rules of the form: $X \rightarrow Y$, where $X$ are features in the design space (e.g., being a LEO architecture) and $Y$ are desired features in the objective space (e.g., being a non-dominated architecture). In this study, we use three metrics to distinguish statistically significant rules, called support, confidence and lift, which are defined as follows:

$$supp(X) = \frac{|\{t \in X \subseteq D\}|}{|D|} \sim P(X) \quad (32)$$

$$conf(X \rightarrow Y) = \frac{supp(X \cap Y)}{supp(X)} \sim P(Y|X) \quad (33)$$

$$lift(X \rightarrow Y) = \frac{supp(X \cap Y)}{supp(X)supp(Y)} \sim \frac{P(X \wedge Y)}{P(X)P(Y)} \quad (34)$$

where $X$ is a binary feature of interest, characterized by the set of designs that have the feature, $t$ is one particular design, and $D$ is the design population.

High support for $X$ means that $X$ appears frequently in the design population. High confidence for the rule $X \rightarrow Y$ means that most designs that have X also have Y (i.e., it is used as a rule strength indicator), and high lift means that there is likely a statistical dependency between $X$ and $Y$. Note that there is generally a trade-off between $conf(X \rightarrow Y)$ and $conf(Y \rightarrow X)$, so a single high confidence by itself is not indicative of good predictive power – a high value for the other confidence or lift do provide more interesting evidence of a "driving feature", i.e., an interesting feature that appears to be driving the results.

A set of baseline (order 1) features was constructed by considering each value that each decision can take: $X_{ij} \equiv (x_i = x_{ij})$ where $x_i$ is the ith decision and $x_{ij}$ is the jth value that the ith decision can take. Then, the a priori algorithm was used to identify the driving features up to order 2– driving features were characterized by a confidence rating $\geq 0.9$ and a lift rating $\geq 1.0$ - in the Pareto front. A priori starts by exploring order 1 features and after eliminating those that do not make user-defined thresholds for support and other interestingness measures, it looks at higher-order features (i.e., conjunctions of selected order 1 features).

### 3) Sensitivity analysis

#### a) Design decisions

Variance-based sensitivity analysis methods are used to understand which decisions $(X_i)$ are driving the variability in the metrics (Y). In particular, Sobol's first order, $S_i$, [58] and total order, $S_i^T$, [59] sensitivity indices are used to study the main effects of each decision on each metric.

$$S_i = \frac{V_{X_i}[E_{X_{-i}}(Y|X_i)]}{V[Y]} \quad (35)$$

$$S_i^T = 1 - \frac{V_{X_{-i}}[E_{X_{-i}}(Y|X_{-i})]}{V[Y]} \quad (36)$$

where $V_{X_i}$ is the data variance observed due to changes in decision $X_i$ alone and $V_{X_{-i}}$ the variance attributed to all decisions but $X_i$, i.e. $X_i$ is considered fixed.

#### b) Model parameters

Five hundred and forty different scenarios were designed in order to introduce variability in key model parameters and assumptions. In particular, the scenarios were composed of all the combinations of the following options: learning factor {85%, 90%}, dry mass variation {40%, -20%, 0%, 20%, 40%}, EOL disposal {BAU, deorbit subsystem], launch cost {-80%,-40%,0%}, user willingness to wait {1, 5, 10}[min.] and running operational costs {0, 2, 4}[$M/sat/year]. Predicated on the ground segment being deployed, running operational costs would consist mainly of mission operations (staffing costs) as well as hardware and software maintenance costs to a lesser extent. For simplicity, the running operational costs are assumed to scale linearly with the number of operational satellites.

### III. RESULTS

In this section, we present the results from tradespace analysis including Pareto analysis, data mining, and sensitivity analysis. In the Pareto Analysis subsection, the tradespace plot is presented followed by a discussion of the visible tradeoffs among the Pareto front architectures. The results correspond to the baseline scenario, which incorporates all the assumptions described in the Methods section. In the Data Mining subsection, the driving features of the Pareto front architectures are identified and interpreted in the context of these underlying assumptions. Finally, we examine the sensitivity of the results to the design decisions and key model parameters such as satellite dry mass, learning factor, EOL disposal strategy, launch costs, user willingness to wait and running operational costs. These results are discussed in the Sensitivity Analysis subsection, which also includes a characterization of robustness to single-satellite failure.

### A. Pareto analysis

The non-dominated sorting algorithm identified 23 architectures on the Pareto front. Table XIII presents the entire set of 23 non-dominated architecture solutions with the corresponding sets of decisions and performance/cost metrics, while Figure 5 shows the tradespace plot including all the architectures (blue) with particular focus on the Pareto front (red) and reference architectures (black). The architectures are divided into four different categories of orbit altitude regimes: LEO [780km and 1250km], Low MEO [8330km and 12525km], MEO [20188km and 23229km] and High MEO [30967km]. In Figure 5, each of these categories is represented by a different color and placed in a separate subplot.

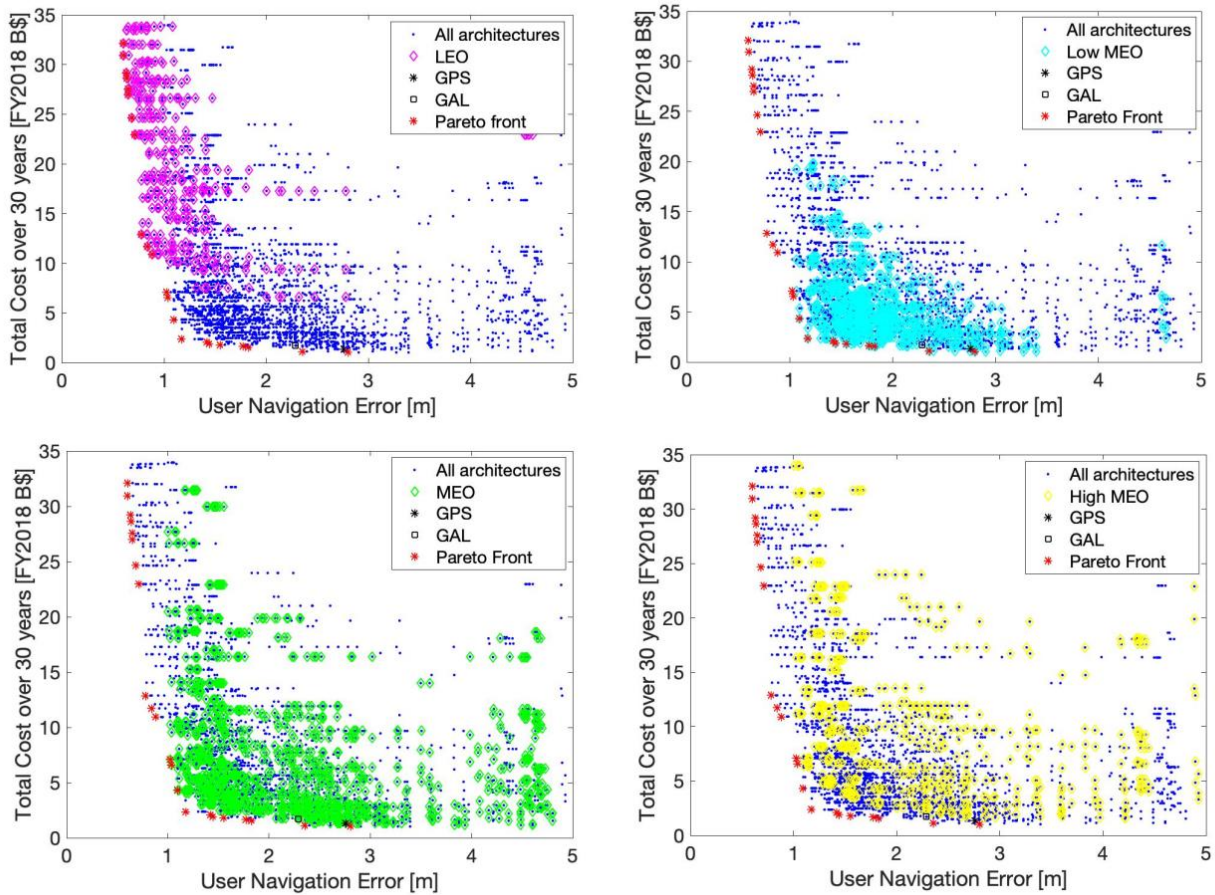

Fig. 5 - Tradespace plot highlighting LEO architectures (upper left), low MEO architectures (upper right), MEO architectures (lower left) and high MEO architectures (lower right).

TABLE XIII
PARETO FRONT ARCHITECTURES IN THE BASELINE CASE ("ELBOW" ARCHITECTURE HIGHLIGHTED)

| ID | Decisions | | | | | | | Metrics | | Additional Information | | | | | | | |
|---|---|---|---|---|---|---|---|---|---|---|---|---|---|---|---|---|---|
| | SV orbit altitude | # SV | Orbit Incl. | # Orbit Planes | Rx Signal Power | Signal Freq. | SV lifetime | NAV Error [m] | Total Cost (30 years) [$B] | S/C Power [W] | Dry Mass [Kg] | Wet Mass [Kg] | Flight Unit Cost [$M] | Launch Costs [$M] | DOP | UERE [m] | Multipath Decorr. Time [min] |
| 1 | 780 | 720 | 87 | 24 | -145 | 3 | 15 | 0.9 | 10.9 | 488.0 | 281.8 | 305.5 | 24.0 | 4656 | 1.4 | 0.6 | 1.4 |
| 2 | 780 | 840 | 87 | 24 | -150 | 3 | 15 | 0.8 | 11.7 | 441.5 | 258.3 | 280.6 | 24.0 | 4656 | 1.3 | 0.7 | 1.4 |
| 3 | 780 | 840 | 87 | 24 | -145 | 3 | 15 | 0.8 | 12.9 | 488.0 | 281.8 | 305.5 | 24.0 | 5820 | 1.3 | 0.6 | 1.4 |
| 4 | 1250 | 600 | 87 | 24 | -145 | 3 | 15 | 0.7 | 23.0 | 532.7 | 788.8 | 832.8 | 49.7 | 11640 | 1.1 | 0.6 | 1.6 |
| 5 | 1250 | 720 | 87 | 24 | -145 | 3 | 15 | 0.7 | 27.0 | 532.7 | 788.8 | 832.8 | 49.7 | 13968 | 1.0 | 0.6 | 1.6 |
| 6 | 1250 | 720 | 87 | 30 | -150 | 3 | 15 | 0.7 | 24.7 | 455.6 | 707.5 | 747.2 | 49.7 | 11640 | 1.0 | 0.7 | 1.6 |
| 7 | 1250 | 720 | 87 | 30 | -145 | 3 | 15 | 0.6 | 27.6 | 532.7 | 788.8 | 832.8 | 49.7 | 14550 | 1.0 | 0.6 | 1.6 |
| 8 | 1250 | 840 | 87 | 24 | -150 | 3 | 15 | 0.6 | 28.6 | 455.6 | 707.5 | 747.2 | 49.7 | 13968 | 0.9 | 0.7 | 1.6 |
| 9 | 1250 | 840 | 87 | 24 | -145 | 3 | 15 | 0.6 | 30.9 | 532.7 | 788.8 | 832.8 | 49.7 | 16296 | 0.9 | 0.6 | 1.6 |
| 10 | 1250 | 840 | 87 | 30 | -150 | 3 | 15 | 0.6 | 29.2 | 455.6 | 707.5 | 747.2 | 49.7 | 14550 | 0.9 | 0.7 | 1.6 |
| 11 | 1250 | 840 | 87 | 30 | -145 | 3 | 15 | 0.6 | 32.1 | 532.7 | 788.8 | 832.8 | 49.7 | 17460 | 0.9 | 0.6 | 1.6 |
| 12 | 12525 | 30 | 56 | 3 | -155 | 3 | 15 | 2.4 | 1.1 | 624.0 | 331.8 | 355.4 | 24.0 | 582 | 2.2 | 1.1 | 10.2 |
| 13 | 12525 | 30 | 64 | 5 | -155 | 3 | 15 | 2.8 | 1.0 | 624.0 | 331.8 | 355.4 | 24.0 | 485 | 2.7 | 1.1 | 10.2 |
| 14 | 12525 | 48 | 56 | 4 | -155 | 3 | 15 | 1.8 | 1.6 | 624.0 | 331.8 | 355.4 | 24.0 | 776 | 1.7 | 1.1 | 10.2 |
| 15 | 12525 | 48 | 64 | 3 | -155 | 3 | 15 | 1.8 | 1.7 | 624.0 | 331.8 | 355.4 | 24.0 | 873 | 1.7 | 1.1 | 10.2 |
| 16 | 12525 | 60 | 64 | 3 | -155 | 3 | 15 | 1.6 | 1.8 | 624.0 | 331.8 | 355.4 | 24.0 | 873 | 1.5 | 1.1 | 10.2 |
| 17 | 12525 | 60 | 64 | 5 | -155 | 3 | 15 | 1.4 | 1.9 | 624.0 | 331.8 | 355.4 | 24.0 | 970 | 1.4 | 1.1 | 10.2 |
| 18 | 12525 | 60 | 64 | 6 | -155 | 3 | 15 | 1.4 | 2.1 | 624.0 | 331.8 | 355.4 | 24.0 | 1164 | 1.3 | 1.1 | 10.2 |
| 19 | 12525 | 84 | 64 | 6 | -155 | 3 | 15 | 1.2 | 2.4 | 624.0 | 331.8 | 355.4 | 24.0 | 1164 | 1.1 | 1.1 | 10.2 |
| 20 | 12525 | 84 | 64 | 6 | -150 | 3 | 15 | 1.1 | 4.3 | 1064.9 | 518.7 | 552.2 | 39.9 | 2328 | 1.1 | 1.0 | 10.2 |
| 21 | 20188 | 24 | 56 | 4 | -155 | 1 | 15 | 12.0 | 0.9 | 617.9 | 314.8 | 333.9 | 24.0 | 388 | 2.9 | 4.2 | 15.9 |
| 22 | 20188 | 84 | 64 | 6 | -150 | 3 | 15 | 1.0 | 6.6 | 1774.8 | 778.3 | 816.2 | 49.7 | 4074 | 1.0 | 1.0 | 15.9 |
| 23 | 23229 | 84 | 64 | 6 | -150 | 3 | 15 | 1.0 | 7.1 | 2122.9 | 915.0 | 955.3 | 59.9 | 4074 | 1.0 | 1.0 | 18.4 |
| GPS | 20188 | 24 | 56 | 6 | -155 | 3 | 15 | 2.8 | 1.3 | 848.4 | 410.6 | 434.0 | 40.0 | 582 | 2.6 | 1.1 | 15.9 |
| GAL | 23229 | 27 | 56 | 3 | -155 | 3 | 15 | 2.3 | 1.7 | 958.5 | 456.7 | 480.8 | 40 | 873 | 2.1 | 1.1 | 18.4 |

From visual inspection, it is clear that Low MEO and MEO architectures dominate most of the Pareto front so it is sensible to assume that most decision-makers would prefer to choose one of these orbit categories. Decision-makers who are willing to pay more than 10 billion dollars to achieve that last bit of performance below 1 meter would favor LEO constellations. Alternatively, it is reasonable to focus on the region located on the 'elbow' of the tradespace plot, which appears to represent a good trade-off among the more affordable (< 6 $B) architectures. An architecture that seems particularly good in this region is highlighted in Table XIII and represents a constellation at 12525km altitude, 64° inclination and 84 satellites in 6 orbital planes. Finally, it is worth noting that the reference architectures are close to being non-dominated; however, due to the sharp form of the Pareto frontier, many architectures closer to the elbow region provide significantly better performance at similar costs.

### B. Data mining

The following conclusions were drawn from the association rule mining results shown in Table XIV. First, the Pareto front is dominated by architectures with the longest single satellite lifetime, i.e., 15 years. This is in part because the effects of technology obsolescence were not considered, which would favor shorter satellite lifetimes. Second, the Pareto front is also dominated by architectures containing a satellite payload capable of transmitting three different NAV signal frequencies. Third, as mentioned above, LEO and "Low MEO" type constellations dominate (twenty out of twenty-three non-dominated architectures) different regions of the Pareto front. LEO constellations are a driving feature in high-power architectures, which result in the best positioning performance both in terms of accuracy and resistance to jamming and spoofing. This is a consequence of the free path loss and the fact that payload power drives the satellite mass and its production costs. Thus, a 10dB satellite TX power increase at 780, 1250, 8330, 12525, 20188, 23229 and 30967 km orbit altitudes results in approximately 6%, 7%, 68%, 124%, 221%, 257% and 355% increase in satellite dry mass respectively. LEO constellations are characterized by large space segment costs (the most affordable being 6.6 $B). If the analysis is focused on architectures whose total space segment cost is under 6 $B (51% of the architectures) then the "Low MEO" constellations are identified as a driving feature. Consequently, these data mining results give support to the findings in the Pareto Analysis

TABLE XIV
ASSOCIATION RULE MINING RESULTS

| Architectures | Features | Supp X | Conf X_Y | Conf Y_X | Lift X_Y |
|---|---|---|---|---|---|
| All | 15 years lifetime | 0.33 | 0.01 | 1.00 | 3.00 |
| | Triple freq. | 0.33 | 0.01 | 0.96 | 2.87 |
| | 15 years + triple freq. | 0.11 | 0.04 | 0.96 | 8.61 |
| High-Power (-145 dBW) | LEO | 0.11 | 0.04 | 1 | 9.05 |
| Affordable ( < 6 $B) | Low MEO | 0.39 | 0.01 | 0.9 | 2.28 |

TABLE XV
SENSITIVITY ANALYSIS - MAIN EFFECTS

| Sobol Indice | | SV orbit altitude | # SV | Orbit Incl. | # Orbit Planes | Rx Signal Power | Signal Freq. | SV lifetime |
|---|---|---|---|---|---|---|---|---|
| User NAV Error [10⁻²] | First | 1.2 | 12.2 | 0.12 | 1.1 | 1.2 | 67.6 | 0.0 |
| | Total | 17.4 | 27.0 | 1.2 | 10.1 | 2.1 | 77.9 | 0.0 |
| Total Cost [10⁻²] | First | 15.9 | 12.4 | 8.4 | 9.4 | 17.7 | 3.7 | 12.4 |
| | Total | 32.8 | 15.9 | 8.7 | 9.9 | 32.7 | 5.2 | 16.6 |

## C. Sensitivity analysis

The results presented in the previous section ignore uncertainty in both decisions and model parameters. This section partially addresses these limitations by presenting the results of a sensitivity analysis. First, the sensitivity of the metrics to the design decisions is calculated. Then, the robustness of results (e.g., the composition of the Pareto front) to the uncertainty in key model parameters is assessed. Together, these analyses help us gain confidence in the validity of the model and reveal new insights about the design space. Finally, the sensitivity of results to single spacecraft failure is considered.

### 1) Sensitivity to design decisions

The results derived from Sobol's first and total indices are summarized in Table XV, and lead to the following observations: (1) The number of signal frequencies decision dominates the user NAV error. This can be explained by the impact attributed to first-order ionospheric delay errors that cannot be mitigated in single-frequency operation but are effectively removed in dual and triple frequency modes. (2) SV lifetime has no impact on the user NAV error, as expected. (3) There are significant interactions –as captured in the difference between Sobol's total effect and first order indices- involving satellite orbit altitude in both performance and cost metrics, which is also expected. (4) The received signal power has surprisingly little impact on the user navigation error. This can be attributed to the fact that the code tracking error is much smaller than the ionospheric error and because the benefits of higher signal power (such as jamming resistance and indoor capability) are not fully captured in the metric as formulated.

### 2) Sensitivity to model parameters

The simulation results under the 540 scenarios, presented in the methods section, revealed several interesting architectures that are non-dominated in the majority of scenarios (Table XVI). These architectures are mostly located at 1250km and 12525km and are characterized by long satellite lifetime and triple frequency.

These findings provide additional support to the observations made in the previous section. The GPS and Galileo reference architectures were dominated in 100% and 80% of the scenarios, respectively, although they are fairly close to the Pareto front in some cases.

More general results in terms of the relative value of the different orbit altitude regimes for different combinations of operational costs, payload power, user willingness to wait and launch costs are shown in Figures 6, 7, 8 and 9 respectively. The variations are made with respect to the baseline scenario. The plots show the percentage of architectures in the fuzzy Pareto front (all architectures with a Pareto ranking of at most 2) in different orbit altitude regimes for different values of the uncertain parameter.

TABLE XVI
ARCHITECTURES SORTED BY ROBUSTNESS (% OF SCENARIOS WHERE ARCHITECTURE IS NON-DOMINATED)

| | Architectures | | | | | | | |
|---|---|---|---|---|---|---|---|---|
| ID | SV orbit altitude | # SV | Orbit Incl. | # Orbit Planes | Rx Signal Power | Signal Freq. | SV lifetime | % of scenarios where architecture is non-dominated |
| 11 | 1250 | 840 | 87 | 30 | -145 | 3 | 15 | 100 |
| 20 | 12525 | 84 | 64 | 6 | -150 | 3 | 15 | 100 |
| 7 | 1250 | 720 | 87 | 30 | -145 | 3 | 15 | 98 |
| 24 | 12525 | 60 | 64 | 6 | -150 | 3 | 15 | 78 |
| 25 | 12525 | 48 | 64 | 6 | -155 | 3 | 15 | 67 |
| 19 | 12525 | 84 | 64 | 6 | -155 | 3 | 15 | 66 |
| 4 | 1250 | 600 | 87 | 24 | -145 | 3 | 15 | 65 |
| 12 | 12525 | 30 | 56 | 3 | -155 | 3 | 15 | 61 |
| 26 | 12525 | 48 | 64 | 6 | -150 | 3 | 15 | 60 |
| 18 | 12525 | 60 | 64 | 6 | -155 | 3 | 15 | 56 |
| 27 | 20188 | 24 | 56 | 3 | -155 | 3 | 15 | 55 |
| 28 | 20188 | 24 | 56 | 3 | -155 | 2 | 15 | 55 |
| 29 | 1250 | 480 | 87 | 24 | -145 | 3 | 15 | 54 |

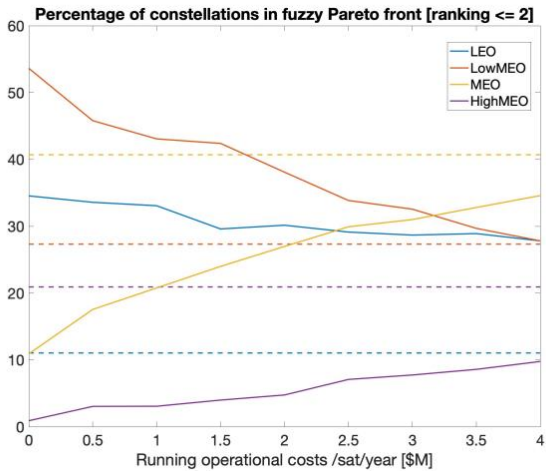

Fig. 6 – Percentage of constellations in fuzzy Pareto front by orbit vs operational costs

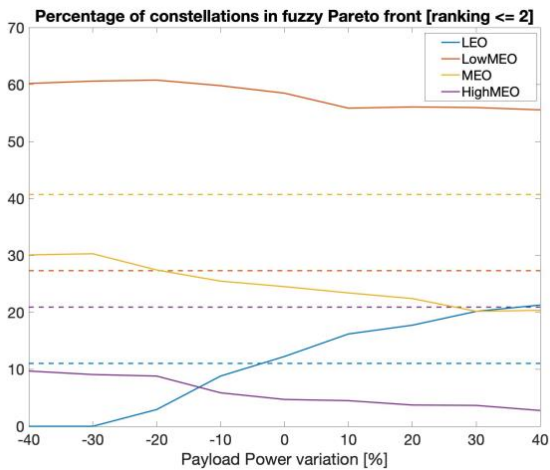

Fig. 7 - Percentage of constellations in fuzzy Pareto front by orbit vs payload power variation

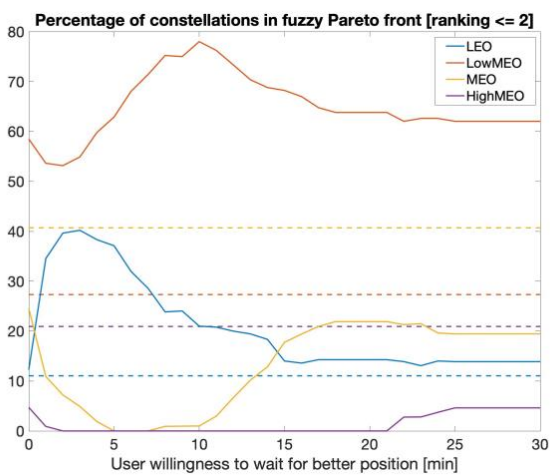

Fig. 8 - Percentage of constellations in fuzzy Pareto front vs user willingness to wait for a better position

For comparison, the dashed lines represent the support of the feature, i.e., the percentage of architectures in those orbit altitudes in the dataset of feasible architectures.

From Figure 6, it is clear that architectures in Low MEO, which contain higher numbers of satellites when compared to MEO, are comparatively less attractive as the operational cost per satellite increases, as expected. However, if operational costs/satellite can be kept under 2 $M/sat/year, these are the dominant architectures. Figure 7 reveals that LEO architectures become more dominant when the payload power requirements increase. This is a consequence of the inverse-square law in radio communications, which effectively translates into a satellite mass penalty as seen before. Figure 8 shows a predominance of LEO and low MEO constellations, which are overly represented in the fuzzy Pareto front when compared with their frequency of occurrence in the dataset (support). Additionally, the Low MEO orbit category is predominant under a use case requiring instantaneous positioning ($T_{wait}$= 0). Finally, regarding Figure 9, we found it more informative to show the performance versus launch cost reduction at all the orbit altitude options as opposed to altitude regimes, since contradicting trends were observed within the Low MEO altitude regime. Specifically, the orbits at 8330km become less attractive when launch costs are reduced, while the 12525km altitude shows the opposite behavior. This gives support to the observation that architectures at 12525km altitude are robust and non-dominated in a wide range of scenarios (as seen in Table XVI), appearing to be good candidate solutions.

### 3) Robustness to satellite failure

As a final sensitivity study, we looked at the impact of satellite failure on GDOP over a time frame that ensures orbit repeatability on the ground. To mitigate the risk of satellite failures, it is common practice to have satellite spares in each orbital plane. Nevertheless, if a satellite fails, the user will be impacted until corrective action takes place. The difference in GDOP values between operational and one-satellite failure modes is thus an indication of architecture robustness. GDOP in failure mode was computed for ten cases of random single satellite failures. The results were found to be consistent across the ten cases with a maximum difference in GDOP of 0.03 (mean 0.005), which means that the main effects are relatively independent of the satellite at fault in cases of single satellite failure mode. This is likely a consequence of the fact that the simulation places the satellites in a perfect Walker Delta pattern, which is rarely observed in practice due to differential orbit perturbation effects. Since this is an expensive calculation, it was only done for a small group of architectures that are representative of different orbit altitude regimes. Figure 10 displays the results. The lines marked with a cross mark represent the $\langle GDOP \rangle_{worst}$ values obtained with one satellite failure, whereas the dot mark represents the values for the whole constellation in operation. As can be seen in the plot, the impact of one satellite failure is global, since it impacts users at all latitudes. As expected, given that the GDOP variation is less pronounced for constellations with a higher number of satellites, we can conclude that these are inherently more robust.

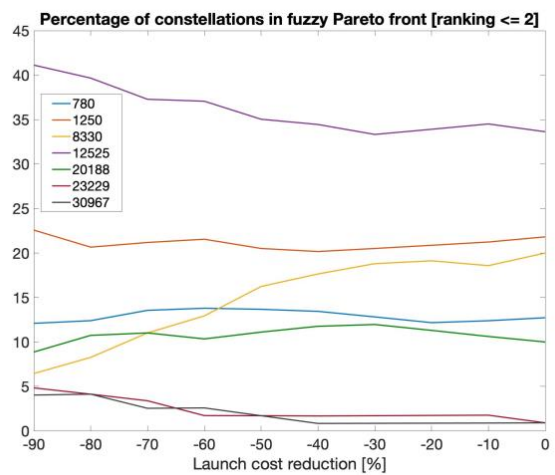

Fig. 9- Percentage of constellations in fuzzy Pareto front by orbit vs launch cost reduction

## IV. CONCLUSION

### A. Contributions

The findings reported here shed new light on a rich tradespace of architectures for next-generation GNSS. The analysis has shown that LEO constellation designs have the potential to achieve the best positioning performance but result in one order of magnitude increase in both the number of satellites and the overall space segment cost when compared to alternative architectures in MEO (see Table XIII). Besides, taking into account the proliferation of space debris in near-polar LEO orbits, these large constellations could prove too risky -e.g. the probability of accidental collision with space objects larger than 10 cm in diameter, throughout the satellite lifetime exceeding 0.1%. In any case, given that the GPS III space segment procurement costs (~200 $M/unit) and launch costs (~100 $M/unit) over 30 years (2 generations) are approximately 20 $B (FY2018) some of the proposed LEO small satellite ($< 500$kg) constellations could be achieved at a similar price point. This is particularly true for the LEO constellations at 780km that benefit from the most favorable space radiation environment (as can be inferred from Table XI). The launch requirements to maintain a LEO constellation are reasonable when considering long satellite lifetimes. As an example, arch. #1 (720 satellites in 24 orbital planes) could be deployed and maintained with only 2 rocket launches per year –each launch populating a different orbital plane with 24 satellites-, assuming a Falcon 9 rocket performance (22,800 kg to LEO), as well as a satellite lifetime and a constellation deployment time of 15 years.

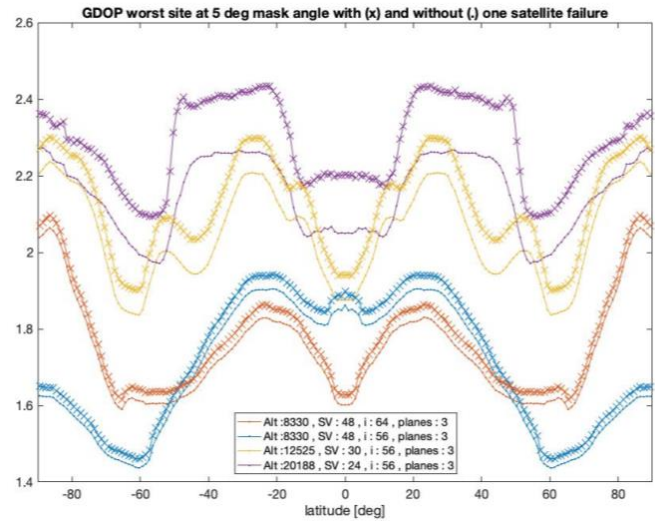

Fig. 10 – Impact of one satellite failure on GDOP worst site @ 5 deg mask angle

Interestingly, we have identified alternative GNSS space segment architectures in Low MEO that have the potential to outperform the current GNSS constellations at similar costs. With respect to the reference architectures in MEO, low MEO architectures have the following advantages: First, the reduced signal power transmission losses allow smaller satellites to achieve the same level of received signal power on the ground. Under the assumptions made in this study, a satellite at 12525km altitude need only transmit a 1kW signal in order to allow for a received signal power on the ground of -150dBW (maximum received signal power of current GPS L5 civil signal), instead of 1.8kW and 2.1kW at GPS and Galileo altitudes respectively. Thus, the benefits of strong signals (such as jamming resistance and code tracking performance in challenging environments) can be accomplished with smaller satellites. Second, stronger Doppler signatures can effectively reduce the impact of multipath errors that are especially relevant in urban environments. Third, based on simulations the required $\Delta V$ for re-entry is 10 to 25% lower, at 12525 and 8330 and km respectively, which is attractive if the BAU deorbiting scenario is deemed too risky and there is a need for a dedicated deorbit subsystem. Fourth, the higher number of satellites in these constellations results in a higher degree of robustness to single-satellite failure.

The architectures at 12525km altitude appear to be particularly robust to variations in key model parameters (dry mass, learning factor, EOL disposal, user willingness to wait and operational costs) as can be seen in Table XVI. One remarkable example is an architecture (ID 19 in Table XIII) defined by a constellation with 84 satellites equally distributed through 6 orbital planes at 64 deg inclination, which was shown to be non-dominated in 100% of the scenarios. For all the stated reasons, the authors believe that architectures in Low MEO – especially at 12525km- outperform existing GNSS at similar costs, particularly in cases where stakeholders' preferences prioritize signal jamming resistance and performance in urban environments. Thus, architectures such as the one highlighted above (ID 19) should be studied further and given serious consideration in the deployment of future GNSS.

### B. Limitations

This study suffered from various limitations. First, it was a screening study that considered only a limited set of possible satellite orbit constellations, even if somewhat representative of a wide orbit altitude range. Second, it focused on modeling the positioning accuracy based on space segment architecture decisions and did not consider other GNSS key performance indicators such as availability, continuity or integrity, since they are highly dependent on the detailed design of the overall system. The results assumed that there is a ground station network capable of tracking and performing state-of-the-art orbit determination. The focus on positioning accuracy based on code processing alone ignored the benefits of carrier phase processing and the impact of signal frequencies choices (combinations) in carrier phase ambiguity resolution. Third, we ignored the potential benefit of shorter lifetimes to enable the incorporation of new technology in every generation. This could partially explain why the longest satellite lifetime was identified as a driving feature of Pareto front architectures for the baseline scenario. Fourth, we did not consider affordability aspects in terms of the yearly budget necessary to deploy the proposed constellations. Finally, this was a static tradespace analysis that did not consider strategic time-dependent aspects of model such as changes in demand or how these constellations would be deployed over time starting from the reference architectures.

### C. Future Work

For further research, we plan to explore the use of optimization algorithms, specifically Multi-Objective Evolutionary Algorithms, to find good solutions over a wider range of design decisions/options including continuous variables. Additional objectives could be added pertaining to the space segment (e.g., geometry repeatability on the ground, DOP geographic variance, constellation stability to orbit perturbations) and the control segment (e.g., NAV message latency, number of tracking stations). Finally, it would be interesting to consider non-Walker Delta constellations, such as hybrid constellations composed of satellites at different orbit altitudes.

### APPENDIX A

Cs/No is computed using the desired received signal power, Cs (Arch. decision #5) and the thermal noise, $N_0$ experienced in a typical GNSS receiver with the following assumptions [60]: (1) Amplifier temperature with noise figure, $\left(N_f\right)_{dB} = 4.3dB$ @ $290K$ : $T_{amp} = 290 \cdot (10^{0.43} - 1) = 490.5K$, (2) Antenna noise temperature, $T_{ant} = 100K$

$$N_0 = 10\,log_{10}\big[k \cdot \left(T_{ant} + T_{amp}\right)\big] = -200.9\ dBW$$
where k = $1.38 \times 10^{-23}\ J.K^{-1}$

For each value of Cs/No, the pseudorange error standard deviation due to tracking noise, $\sigma_{tn}$ [m], was computed for a typical noncoherent discriminator (early-minus-late power), based on Equation 37 below [60] and the following assumptions: (1) BPSK modulation with spreading code rate, $R_c = 1.023\ MHz$ (GPS C/A code), (2) code loop noise bandwidth, $B_n = 1\ Hz$, (3) Early-to-late correlator spacing, $D = 0.1$, (4) pre-detection integration time, $T_{int} = 20\ ms$, (5) double-sided front-end bandwidth, $B_{fe} = 8\ MHz$.

$$\sigma_{tn} \approx \sqrt{\frac{B_n}{2\left(^{c_s}/_{n_0}\right)}\left(\frac{1}{B_{fe}T_c} + \frac{B_{fe}T_c}{\pi - 1}\left(D - \frac{1}{B_{fe}T_c}\right)^2\right)} \qquad (37)$$
$$\cdot \sqrt{\left[1 + \frac{2}{T_{int}\left(^{c_s}/_{n_0}\right) \cdot (2-D)}\right]} \cdot \frac{c}{R_c}$$

where $^{c_s}/_{n_0}$ is the carrier to noise density ratio in linear scale and $T_c$ is the chip period ($T_c = 1/R_c$).

### APPENDIX B

The solar array mass, $m_{SA}$ and the power production at BOL, $P_{BOL}$ is computed as follows [39]:

1. Compute maximum eclipse duration $T_e^{max}[s]$ from orbit altitude, $h_s[km]$, Earth radius, $r_e$ [km] and gravitational constant, $\mu$ [$km^3/s^2$]. These equations provide very similar results to actual eclipse durations derived from GPS data processing.

$$T[s] = 2\pi\sqrt{(r_e + h_s)^3/\mu} \qquad (38)$$

$$T_e^{max}[s] = T \cdot \sin^{-1}(r_e/(r_e + h_s)) /3 \qquad (39)$$

2. Compute power requirements for the solar array during daylight time $T_d$, to power the satellite for the entire orbit, assuming energetic efficiencies in eclipse $X_e$ and daylight $X_d$ of 0.6 and 0.8 respectively and equal power consumption during daylight and eclipse, $P_d = P_e = P_{SC}$

$$P_{SA}[W] = \frac{\left(\frac{P_e T_e}{X_e} + \frac{P_d T_d}{X_d}\right)}{T_d}\ , T_d = T - T_e \qquad (40)$$

3. Determine beginning-of-life (BOL) power production capability, $P_{BOL}^{area}$ per unit area of the array

4. based on the solar constant $k_{sun} = 1368 \ W/m^2$ and assuming GaAs triple-junction cell efficiency, $\eta_{cell} = 28\%$ [61], a maximum sun incidence angle, $\theta = 5°$ and an inherent degradation of the solar array w.r.t. individual cells, $I_d = 78\%$:

$$P_{BOL}^{area}[W/m^2] = \eta_{cell} \cdot k_{sun} \cdot I_d \cdot \cos\theta \qquad (41)$$

5. Compute end-of-life (EOL) power production capability, $P_{EOL}$, taking into account the satellite lifetime, $S_{life}$ and assuming a solar array degradation, $D = 3.7\%$ per year (GPS II-R satellites experienced a 37% degradation over 10 years [62]):

$$P_{BOL}^{area}[W/m^2] = P_{BOL}^{area} \cdot (1 - D)^{S_{life}} \qquad (42)$$

6. Compute the solar array area, $A_{SA}$ and solar array mass, $m_{SA}$ assuming a nominal rigid panel array performance, $Perf_{SA} = 40 \ W/kg$ and compute total power production at BOL:

$$A_{SA}[m^2] = P_{SA}/P_{EOL}^{area} \qquad (43)$$

$$m_{SA}[kg] = P_{SA}/Perf_{SA} \qquad (44)$$

$$P_{BOL}[W] = P_{BOL}^{area} \cdot A_{SA} \qquad (45)$$

APPENDIX C

The transmitted power $P_T$ is calculated through a simple link budget. The link budget is done using as input variables the architecture's target received signal power ($P_R$) and the following variables: (1) maximum satellite-user range, $r_{s2r}$ @ minimum elevation angle ($\eta^{min}$) = 5°, (2) antenna polarization loss, $L_{ant} = 2.0 \ dB$, (3) excess loss (beyond free-space loss), $L_{ex} = 0.5 \ dB$, (4) receiver antenna gain, $G_R = 0 \ dBi$, (5) transmit antenna gain, $G_T = 13 \ dBi$. The link budget equation expressed in terms of the received signal power is given by [63]:

$$P_R = P_T + G_T + G_R - L_{ex} - L_{ant} \qquad (46)$$
$$+ 20\log_{10}(c/(4\pi \cdot f \cdot r_{s2r}^{max}))$$

which depends on the maximum satellite-user range, $r_{s2r}^{max}$ computed using the law of cosines as described in [63], where $r_e$ [km] represents the earth radius:

$$r_{s2r}^{max} \qquad (47)$$
$$= -r_e\sin(\eta^{min}) + \sqrt{(r_e + h_s)^2 - r_e^2\cos^2(\eta^{min})}$$

The corresponding transmitted power level, $P_T[dBW]$ was obtained as follows:

$$P_T = \sum_{i=1}^{k}(P_R - G_R - G_T + L_{ex} + L_{ant} \qquad (48)$$
$$- 20\log_{10}(c/(4\pi \cdot f_i \cdot r_{s2r}^{max})))$$

where $f_1 = L_1 = 1575.42 \ MHz$,
$f_2 = L_2 = 1227.60 \ MHz$,
$f_3 = L_5 = 1176.45 \ MHz$

Where $k$ is equal to 1, 2 or 3 if the architecture requires single, dual or triple signal frequency transmission respectively.

http://www.russianspaceweb.com/o3b-flight4.html.

**BIOGRAPHY**

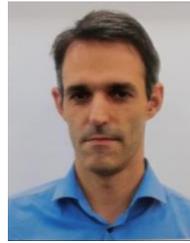

***Filipe Pereira*** *is a Ph.D. candidate in Systems Engineering at Cornell University since 2017. He received the M.Eng. degree in Aerospace Engineering from Instituto Superior Técnico, University of Lisbon, Portugal, in 2002. After his graduation, he worked at the Payload Systems Laboratory at ESA/ESTEC in support of the GALILEO and EGNOS programs. From 2006 to 2014 he worked at ESA/ESOC, where he joined ROSETTA's flight control team and later worked as a Navigation Engineer at the Navigation Support Office. From 2015 to 2017, he was the Comms lead for the Cislunar Explorers mission (NASA's CubeQuest Challenge) at Cornell.*

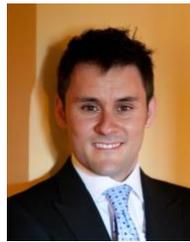

***Daniel Selva*** *is an Assistant Professor of Aerospace Engineering at Texas A&M University, where he directs the Systems Engineering, Architecture, and Knowledge (SEAK) Lab. His research interests focus on the application of knowledge engineering, global optimization and machine learning techniques to systems engineering and architecture, with a strong focus on space systems. Before doing his PhD in Space Systems at MIT, Daniel worked for four years in Kourou (French Guiana) as an avionics specialist within the Ariane 5 Launch team. Daniel has a dual background in electrical engineering and aerospace engineering, with degrees from MIT, Universitat Politecnica de Catalunya in Barcelona, Spain, and Supaero in Toulouse, France. He is a member of the AIAA Intelligent Systems Technical Committee, and in 2018 he was appointed to the European Space Agency's Advisory Committee for Earth Observation.*